\documentclass[twocolumn,final,epj]{svjour}
\usepackage[utf8]{inputenc}
\usepackage[T1]{fontenc}
\usepackage{graphicx}  
\usepackage{amssymb}
\usepackage{amsmath}
\usepackage{hyperref}
\hypersetup{
    colorlinks=true,       
    linkcolor=blue,          
    citecolor=blue,        
    filecolor=magenta,      
    urlcolor=cyan           
}
\usepackage{color}

\begin{document}

\title{Thermoelectric properties of Wigner crystal \\
in two-dimensional periodic potential}

\author{
Mikhail Y. Zakharov\inst{1} 
\and
Denis  Demidov\inst{2}
\and
Dima L. Shepelyansky\inst{3}}

\institute{
Institute of Physics, Department of General Physics, 
Kazan Federal University, 42011 Kazan, Russia
\and
Kazan Branch of Joint Supercomputer Center, 
Scientific Research Institute of System Analysis
\and
Laboratoire de Physique Th\'eorique, IRSAMC, 
Universit\'e de Toulouse, CNRS, UPS, 31062 Toulouse, France
}

\titlerunning{Thermoelectric properties of Wigner crystal in 2D lattice}
\authorrunning{M.Y.~Zakharov, D.~Demidov and D.L.~Shepelyansky}

\date{Dated: 28 October 2019}

\abstract{We study  numerically
transport and  thermoelectric properties of electrons 
placed in a two-dimensional (2D) periodic potential.
 Our results show that
the transition from sliding to pinned phase
takes place at a certain critical amplitude of lattice potential 
being similar to the Aubry transition for the 
one-dimensional Frenkel-Kontorova model.
We show that the 2D Aubry pinned phase is characterized by 
high values of Seebeck coefficient  
$S \approx 12$.
At the same time we find that the value of
Seebeck coefficient is significantly influenced by the
geometry of periodic potential.
We discuss possibilities to test the 
properties of 2D Aubry phase with 
electrons on a surface of liquid helium.
}


\maketitle

\section{Introduction}
\label{sec:1}

The Wigner crystal \cite{wigner}  
has been realized with a variety of solid-state
systems including
electrons on a surface of liquid helium \cite{konobook} and
quantum wires in solid state systems 
(see e.g. review \cite{matveev}).
For one-dimensional (1D) case 
it was theoretically shown that 
the properties of  Wigner crystal
in a periodic potential are highly nontrivial and interesting \cite{fki2007}.
At a weak amplitude of periodic potential
the Wigner crystal slides freely 
while above a critical amplitude of potential
it is pinned by a periodic lattice.

It was shown \cite{fki2007} that this system can be approximately reduced to 
the Frenkel-Kontorova model (see detailed description in \cite{obraun})
corresponding to a chain of particles connected by linear springs
and placed in a periodic potential. In the Frenkel-Kontorova model
the equilibrium positions of particles are described by the
Chirikov standard map \cite{chirikov} which
represents a cornerstone model of area-preserving maps
and dynamical chaos (see e.g. \cite{lichtenberg,meiss}).
It is known that this map describes a variety of physical systems
\cite{stmapscholar}. 
A small potential amplitude corresponds to 
a small kick amplitude of the Chirikov standard map
and in this regime the phase space is covered by isolating
Kolmogorov-Arnold-Moser (KAM) invariant curves.
The rotation phase frequency of a KAM curve
corresponds to a fixed 
irrational density $\nu$ of particles per period.
In this KAM regime the spectrum of small oscillations of
particles near their equilibrium
positions is characterized by a linear phonon (or plasmon) 
spectrum similar to those in a crystal.
Thus in the KAM phase a chain can slide freely in space.
However, for a potential amplitude above a certain critical value
the chain of particles is pinned by the lattice
and the spectrum of oscillations has an optical gap
related to the Lyapunov exponent of the invariant cantori
which replaces the KAM curve. The appearance of this phase had been
rigorously shown by Aubry \cite{aubry} and 
is known as the Aubry pinned phase.
In \cite{fki2007} it is shown that for charged particles
with Coulomb interactions the charge positions
are approximately described by the Chirikov standard map
and that the transport of Wigner crystal in a periodic potential
is also characterized by a transition from the sliding KAM phase
to the Aubry pinned phase.

A new reason of interest to a Wigner crystal transport 
in a periodic potential
is related to the recent results showing that the Aubry
phase is characterized by very good thermoelectric properties with
high Seebeck coefficient $S$ and high figure of merit 
$ZT$ \cite{ztzs,lagesepjd}. 
The fundamental aspects of thermoelectricity had been 
established in far 1957 by Ioffe \cite{ioffe1,ioffe2}.
The thermoelectricity of a system is characterized by 
the Seebeck coefficient $S=-\Delta V /\Delta T$
(or thermopower). It is expressed through a voltage difference $\Delta V$
compensated by a temperature difference $\Delta T$.
Below we use units with a charge $e=1$ and the Boltzmann constant $k_B=1$ 
so that $S$ is dimensionless ($S=1$ corresponds to
$S \approx 88 \rm\mu V/K$ (microvolt per Kelvin)).
The thermoelectric materials are ranked by a figure of merit 
$ZT=S^2\sigma T/ \kappa$ \cite{ioffe1,ioffe2}
with $\sigma$ being an electric conductivity,
$T$ being a temperature and $\kappa$ being the thermal conductivity
of material.

Nowadays the needs of efficient energy usage stimulated extensive 
investigations of various materials with high
characteristics of thermoelectricity as reviewed in
\cite{sci2004,thermobook,baowenli,phystod,ztsci2017}.
The aim is to design materials with $ZT >3$
that would allow an efficient conversion 
between electrical and thermal forms of energy.
The best thermoelectric materials created till now have $ZT \approx 2.6$.
At the same time the numerical modeling reported 
for a Wigner crystal reached values $ZT \approx 8$ \cite{ztzs,lagesepjd}. 
However, these results are obtained in 1D case while 
the thermoelectric properties of  Wigner crystal
in a two-dimensional (2D) periodic potential
have not been studied yet. Also the physics of the Aubry transition
in 2D has not been investigated in detail.
It has been argued \cite{dresselhaus} that high thermoelectric properties 
should appear in low-dimensional systems
and thus the studies of 2D case and its comparison
with 1D one are especially interesting.

As possible experimental systems with a Wigner crystal in a periodic potential
we point to electrons on liquid helium \cite{konobook}.
The experimental investigations of such systems have been already 
started with electrons on liquid helium with a quasi-1d channel 
\cite{kono1d} and  
with a periodic 1D or 2D potential \cite{konstantinov}.
Another physical system is represented
by cold ions in a periodic 1D potential
proposed in \cite{fki2007}. In this field
the proposal \cite{fki2007}
attracted the interest of experimental groups
with first results reported in
\cite{haffner2011,vuletic2015sci}. 
Later the signatures of
the Aubry-like transition have been reported 
by the  Vuletic group with 
5 ions \cite{vuletic2016natmat}.
The chains with a larger number of ions are now 
under investigations in \cite{ions2017natcom,drewsen}. 
However, at present it seems rather difficult to extend cold
ions experiments to 2D case.
Thus we expect that the most promising experimental
studies of thermoelectricity of Wigner crystal in 2D
periodic potential should be the extension of experimental
setups with electrons on liquid helium
reported in \cite{kono1d,konstantinov}.
It is also possible that other physical systems like 
two-dimensional colloidal monolayers, where the observation of Aubry transition
has been reported recently \cite{bechingerprx},
can open complementary possibilities
for experimental modeling of thermoelectricity.

In this work we present the numerical study
of transport and thermolectric properties 
of Wigner crystal in 2D lattice.
We use the numerical vector codes reported in \cite{zakharovprb} 
which employ GPGPU computers thus allowing
to simulate numerically the dynamics of a large number
of electrons. We present the results for the crystal velocity
and Seebeck coefficient at different system parameters 
and different lattice configurations.

The paper is composed as follows: the model description
is given in Section 2, the equations for equilibrium charge 
positions are discussed in Section 3,
properties of electron current are analyzed in Section 4,
the results for Seebeck coefficient at different
lattice geometries are presented in Section 5
and the discussion is given in Section 6.

\section{Model description}
\label{sec:2}

The Hamiltonian of a chain of  charges in a 2D periodic potential
has the form
\begin{eqnarray}
\nonumber
H &=& {\sum_{i=1}^{N_{tot}}} \big( \frac{{P_{ix}}^2}{2} + \frac{{P_{iy}}^2}{2} 
   + V(x_i,y_i) \big) + U_C \; ,\\
\nonumber
 U_C &=& \sum_{i > j} \frac{ 1} {\sqrt{(x_i - x_j)^2+(y_i-y_j)^2+a^2}} \; ,\\
\nonumber
V=V_1(x_i,y_i) &=& - K \big( \cos x_i +  \cos y_i \big) \; ;\\
\nonumber
V=V_2(x_i,y_i) &=& - K \big( \cos (x_i +y_i/2) +  \cos y_i \big) \; ;\\
\nonumber
V=V_3(x_i,y_i) &=& - K \cos x_i  - 0.5 K_h  \big( y_i - h/2)\big)^2 \; ,\\
\label{eq:ham2}
\end{eqnarray}
where  2D momenta $P_{ix}, P_{iy}$ are conjugated to particle  
space coordinates $x_i , y_i$ and $V(x_i,y_i)$ is an external potential.
We consider two geometries of periodic potential
with square $(V=V_1)$ and diamond $(V=V_2)$ lattices.
In addition we consider a channel model  $(V=V_3)$
with a periodic lattice along $x$-axis and oscillator confinement 
in $y$-axis.
The Hamiltonian
is written in dimensionless units
where the lattice period is $\ell=2\pi$
and particle mass and charge are $m=e=1$.
In these atomic-type units 
the system parameters are measured in physical
units:   $r_a= \ell/2\pi$ for length,
$\epsilon_a = e^2/r_a = 2\pi e^2/\ell$ for energy,
$E_{adc} = \epsilon_a/e r_a$  
for applied static electric field,
$v_a=\sqrt{\epsilon_a/m}$ for particle velocity $v$,
$t_a =  e r_a \sqrt{m/\epsilon_a}$ for time $t$.
The temperature $T$ (or $k_B T$)
is also measured in this dimensionless units,
thus for $\ell = 1 \mu m$ the dimensionless temperature
$T=0.01$ corresponds to the physical temperature
$T= 0.01 \epsilon_a/k_B  = 0.02 \pi e^2/(\ell k_B) 
\approx 1 \rm K $ (Kelvin).

As in \cite{ztzs,zakharovprb} the electron dynamics 
is modeled in the frame of Langevin approach (see e.g. \cite{politi})
described by  equations of motion:
\begin{eqnarray}
\nonumber
\dot{P}_{ix} & = & \dot{v}_{ix}= -\partial H/\partial x_i +E_{dc} -\eta P_{ix}+g \xi_{ix}(t) \;, \\
\nonumber
\dot{P}_{iy} & = &\dot{v}_{iy}= -\partial H/\partial y_i -\eta P_{iy}+g \xi_{iy}(t) \;, \\ 
\dot{x_i} & = & P_{ix}  = v_{ix} \; ,  \dot{y_i} = P_{iy}  = v_{iy} \; .
\label{eq:langevin}
\end{eqnarray}
The parameter $\eta$ phenomenologically describes 
dissipative relaxation processes, and 
the amplitude of Langevin force $g_L$ is given 
by the fluctuation-dissipation theorem $g_L=\sqrt{2\eta T}$
where $T$ is the system temperature.
Here we also use particle velocities $v_{ix}=P_{ix} , v_{iy}=P_{iy}$ (since mass is unity).
As usual, the normally distributed random variables $\xi_i$ are 
defined by correlators
$\langle\langle\xi_i(t)\rangle\rangle=0$,
$\langle\langle\xi_i(t) \xi_j(t')\rangle\rangle=\delta_{i j}\delta(t-t')$.
The amplitude of the static force, or electric field,  is given by $E_{dc}$.

The equations (\ref{eq:langevin}) are solved numerically
with a time step $\Delta t$, at
each such a step the Langevin contribution is taken into
account, As in \cite{zakharovprb} we usually use $\Delta t = 0.02$ 
and $\eta=0.1$ with the results being not sensitive to these parameters.
The length of the system in $x-$axis is taken to
be $2\pi L$ 
with $L$ being the integer number of periods.
In $y$-axis we use $L_y$ periodic cells 
with periodic boundary conditions.
In $x$-direction we consider the motion on a ring with 
a periodic boundary conditions
or an elastic wall placed at $x=0$ (to have balanced charge
interactions).
There are $N_{tot}$ electrons in $L L_y$ cells and
the dimensionless charge density is $\nu_2 =N_{tot}/(LL_y)$.
We use $N_{tot}=L_y N$ so that we have 1D density in each of
$L_y$ stripes being $\nu=\nu_2=N/L$. Thus for the Fibonacci values of $N=21,34,55...$ 
and $L = 13,21,34...$ we have $\nu \approx 1.618$ corresponding to 1D case
studied mainly in \cite{fki2007,ztzs}.
The numerical simulations are performed up to dimensionless times
$t=2 \times 10^6$ at which the system is in the steady-state.

\begin{figure}[t]
\begin{center}
\includegraphics[width=0.48\textwidth]{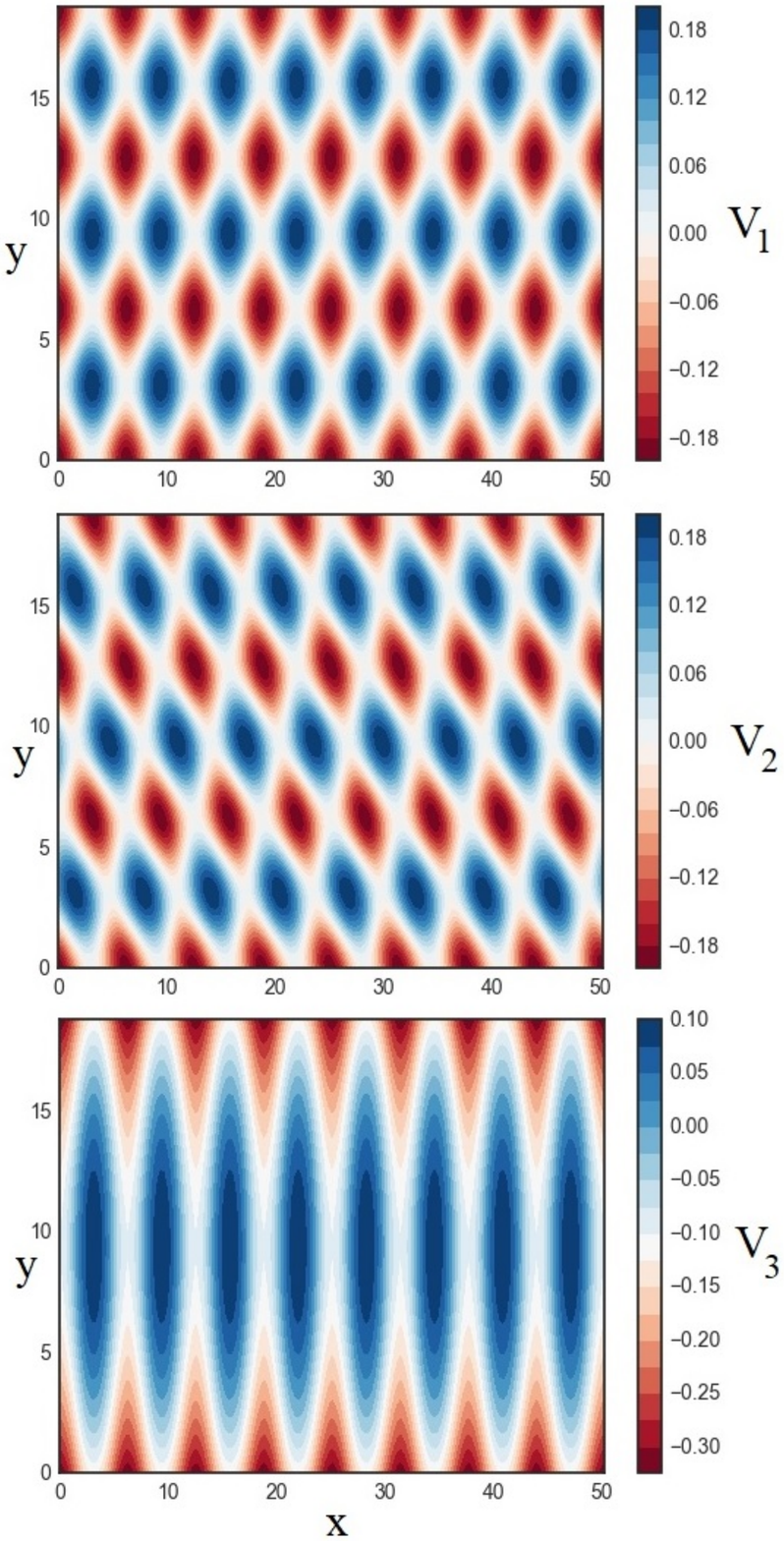}
\end{center}
\caption{\label{fig1} 
Three potentials shown by color
for the three cases from (\ref{eq:ham2})
at $V=V_1$ (top panel),
$V=V_2$ (middle panel),
$V=V_3$ (bottom panel, here $K=0.1$, $K_h=0.005$).
 }
\end{figure}

As in \cite{zakharovprb} the numerical simulations are based on the combination of
Boost.odeint~\cite{odeint} and VexCL~\cite{demidov2013,vexcl} libraries and
employ the approach described in \cite{ahnert2014} in order to accelerate the
solution with NVIDIA CUDA technology.  The equations \eqref{eq:langevin}   are solved
by Verlet method, where
each particle is handled by a single GPU thread.  Since Coulomb interactions
are decreasing with distance between particles, 
the interactions for the 2D case are cut off at the
radius $R_C=6\ell = 12\pi$, that allows to reduce the computational complexity of the
algorithm from $O(N^2)$ to $O(N \log N)$.  
In order to avoid close encounters
between particles leading to numerical instability, the screening length
$a=0.7$ is used.  At such a value of $a$ the interaction energy is still
significantly larger than the typical kinetic energies of particles ($T \ll 1/a$) 
and the screening does not significantly affect the interactions of
particles. 
The source code  is available at \\
https://gitlab.com/ddemidov/thermoelectric2d .  The numerical
simulations were run at OLYMPE CALMIP cluster \cite{olympe} with NVIDIA Tesla
V100 GPUs and partially at Kazan Federal University with NVIDIA Tesla C2070
GPUs. 

\begin{figure}[t]
\begin{center}
\includegraphics[width=0.48\textwidth]{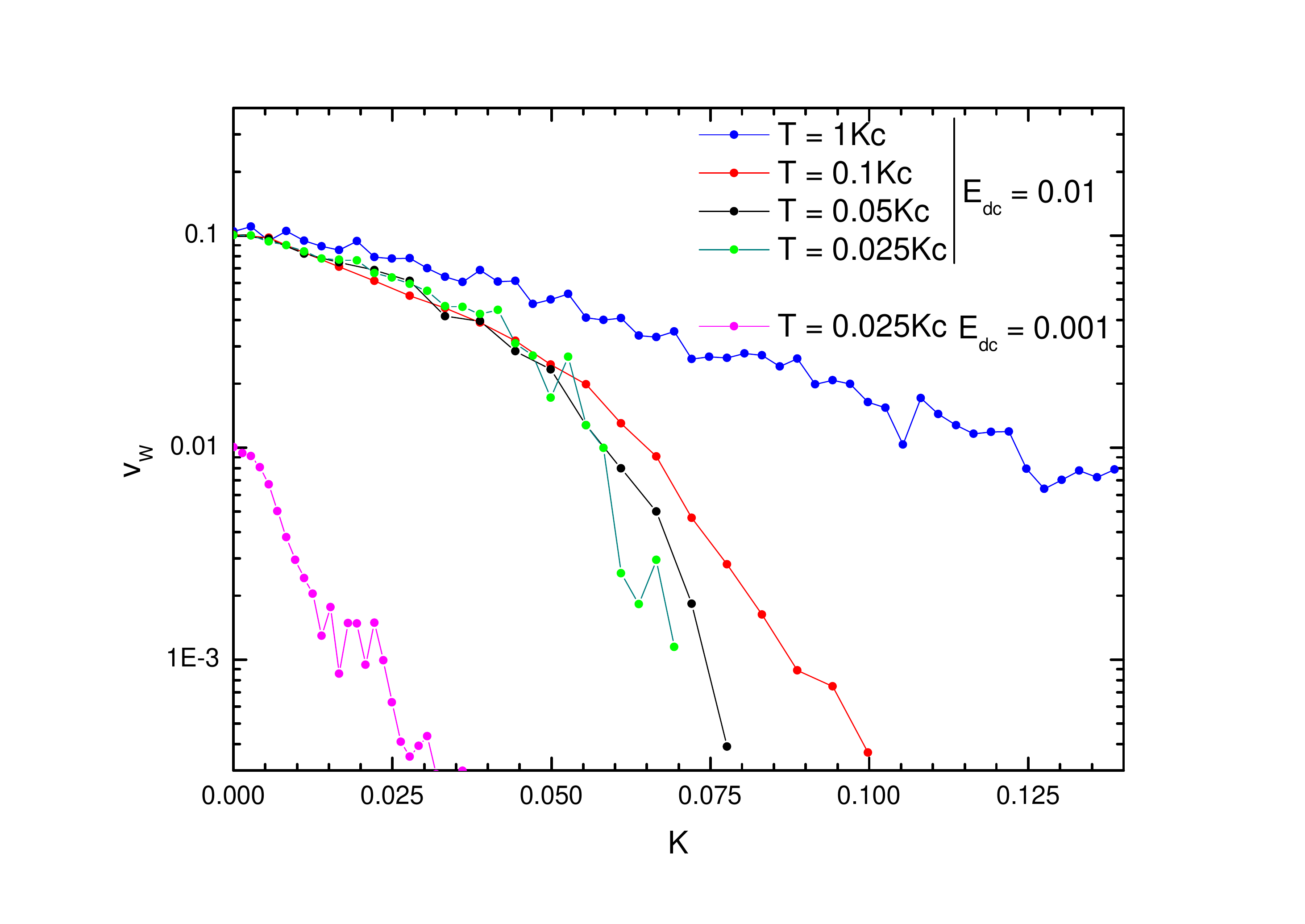}
\end{center}
\caption{\label{fig2} 
Dependence of the Wigner crystal (5 stripes) $v_W$ on 
the potential parameter $K$ for 2D system with $\nu = 34/21$ 
at different values of driving force $E_{dc}$ and temperature $T$;
here $K_c=0.0462$ is the critical potential amplitude
for the Aubry transition in 1D case. The system has $L=21$ cells
in $x$-axis and $L_y=5$ cells in $y$-axis and $N=34 \times 5$ electrons;
here the potential $V=V_1$ in (\ref{eq:ham2}).
 }
\end{figure}

In this work we consider only the problem of classical
charges. Indeed, as shown in \cite{fki2007} the dimensionless Planck
constant of the system is 
$\hbar_{\rm eff}=$ $\hbar/$ $(e \sqrt{m \ell/2\pi})$.
For a typical lattice period $\ell \approx 1 \rm\mu m  $, $\nu \sim 1$ 
and  electrons on a periodic potential of liquid helium
we have a very small effective Planck
constant $\hbar_{\rm eff} \approx 2 \times 10^{-3}$.

\section{Equilibrium positions of electrons}
\label{sec:3}

As for the 1D case
the equilibrium static positions of electrons in a periodic potential
are determined by the conditions $\partial H/ \partial x_i = \partial H/ \partial y_i  =0$, 
$P_{ix}=P_{iy}=0$ \cite{fki2007,aubry}.
In the approximation of nearest  neighbor
interacting electrons, taking into account only 
nearby cells in $x$ and $y$ directions, 
this leads to the  map for recurrent
electron  positions $x_i, y_j$
\begin{eqnarray}
\nonumber
p_{x,i+1} = p_{x,i} + K g_x(x_i) \; , \; \; x_{i+1} = x_i+ F_{p_{x,i+1}} \; ,\\
p_{y,j+1} = p_{y,j} + K g_y(y_j) \; , \; \; y_{j+1} = y_j+ F_{p_{y,j+1}} \; ,
\label{eq:map}
\end{eqnarray}
where the effective momentum conjugated to $x_i$ and $y_j$, are
$p_{x,i} = - d V/d x_i =(x_i-x_{i-1})/R^3$ and 
$p_{y,j} = - d V/d y_j =(y_j-y_{j-1})/R^3$ 
with $R^2 = (x_i-x_{i-1})^2 + (y_j-y_{j-1})^2 +a^2$
and the kick functions
$K g_x(x_i)= \left.-dV/dx\right|_{x=x_i}$ $ = - K \sin x_i$ and
$K g_y(y_j)= \left.-dV/dy\right|_{y=y_j}$ $ = - K \sin y_j$ (for $V=V_1$).
The functions $F_{p_{x,i+1}} ; F_{p_{y,j+1}}$ express the changes 
$ x_{i+1} - x_i ; y_{j+1} - y_j$.
For 1D case the recursive map for electron positions has an explicit
symplectic form \cite{fki2007,lagesepjd,zakharovprb} (e.g. see Eq.(3) in \cite{lagesepjd}).
This 1D map can be approximately reduced to the Chirikov standard map 
\cite{chirikov,fki2007,lagesepjd} that allows to obtain the analytical
dependence for the Aubry transition on charge density.
We note that if we neglect electron interactions between different stripes
then we obtain approximately from (\ref{eq:map}) the 1D map studied in 
\cite{fki2007,lagesepjd,zakharovprb}.

However, interactions between stripes, represented by cells in $y$-axis, 
play an important role and
hence in 2D case the map is much more complicated having
an implicit form. Also from the dynamical view point it corresponds to the case
of two times represented by indices $i$ and $j$.
Such maps have been never studied from a mathematical view point
that makes their analysis very complicated.

Due to these reasons we do not enter in the mathematical
analysis of such maps. Instead, we directly study
the transport properties on electrons 
described in next Sections. 
 
\section{Properties of electron current}
\label{sec:4}

In the frame of described Langevin approach we determine numerically the average flow 
velocity $v_W$ of the Wigner crystal in $x$-direction under the
influence of a static electric field $E_{dc}$ using periodic boundary conditions
in $x$-axis.
In absence of periodic potential the crystal flows with 
the free electron velocity $v_0 = E_{dc}/\eta$ 
(such a case was also discussed in \cite{zakharovprb}).

A typical dependence of $v_W$ on 
the potential amplitude $K$ at different values of 
temperature $T$ and static field $E_{dc}$ are shown in Figure~\ref{fig2}
for the potential $V=V_1$ in (\ref{eq:ham2}).
This data shows that at fixed $T$ the current velocity
$v_W$ drops exponentially with increase of the potential
amplitude $K$.  This is consistent with the presence of
Aubry transition from the Aubry pinned phase at $K>K_{c2}$ to
the KAM sliding phase at $K < K_{c2}$.
Here $K_{c2}$ is a certain critical amplitude
of the transition. We can estimate that
$K_{c2} \sim 0.02$ being approximately by a factor of
2 smaller comparing to the critical amplitude $K=K_c=0.0462$
in 1D at $\nu = 1.618..$ \cite{fki2007,lagesepjd,ztzs}.
At the same time an exact determination of $K_{c2}$
requires a detailed numerical analysis 
of transport at rather small $E_{dc}$ values
and small temperatures.
Indeed the comparisons of $v_W$ values at
$E_{dc}= 0.01$ and $0.001$ shows that at small $K$ values
we have a linear regime with $v_W \sim E_{dc}/\eta$
but at $K \approx 0.02$ such a linear response starts to be
destroyed pointing that $K_{c2}$ can be somewhat smaller
with $K_{c2} \sim 0.015$.
In fact, the situation in 2D case
is more complicated compared to 1D case.
Indeed, in 1D for $K> K_c$ there are no
KAM curves and electrons should overcome
a potential barrier to propagate along the lattice
(while for $K<K_c$ they can freely slide along the lattice
as it is guaranteed by the Aubry theorem \cite{aubry}).
In 2D case the situation is more complex since even 
at large $K> K_{c2}$ there are formally 
straight paths propagating in $x$-direction,
but it is possible that they are not really accessible due to interactions
between electrons. Thus we estimate that in 2D lattice
with $V=V_1$ in (\ref{eq:ham2}) we have at $\nu \approx 1.618$
the Aubry transition at $K_{c2} \approx 0.015 - 0.02$.
The exact value of $K_{c2}$ is not very
important for our further
thermoelectric studies which are
performed at $K$ values being significantly larger
then $K_{c2}$ and at larger temperatures $T$. 

\begin{figure}[t]
\begin{center}
\includegraphics[width=0.23\textwidth]{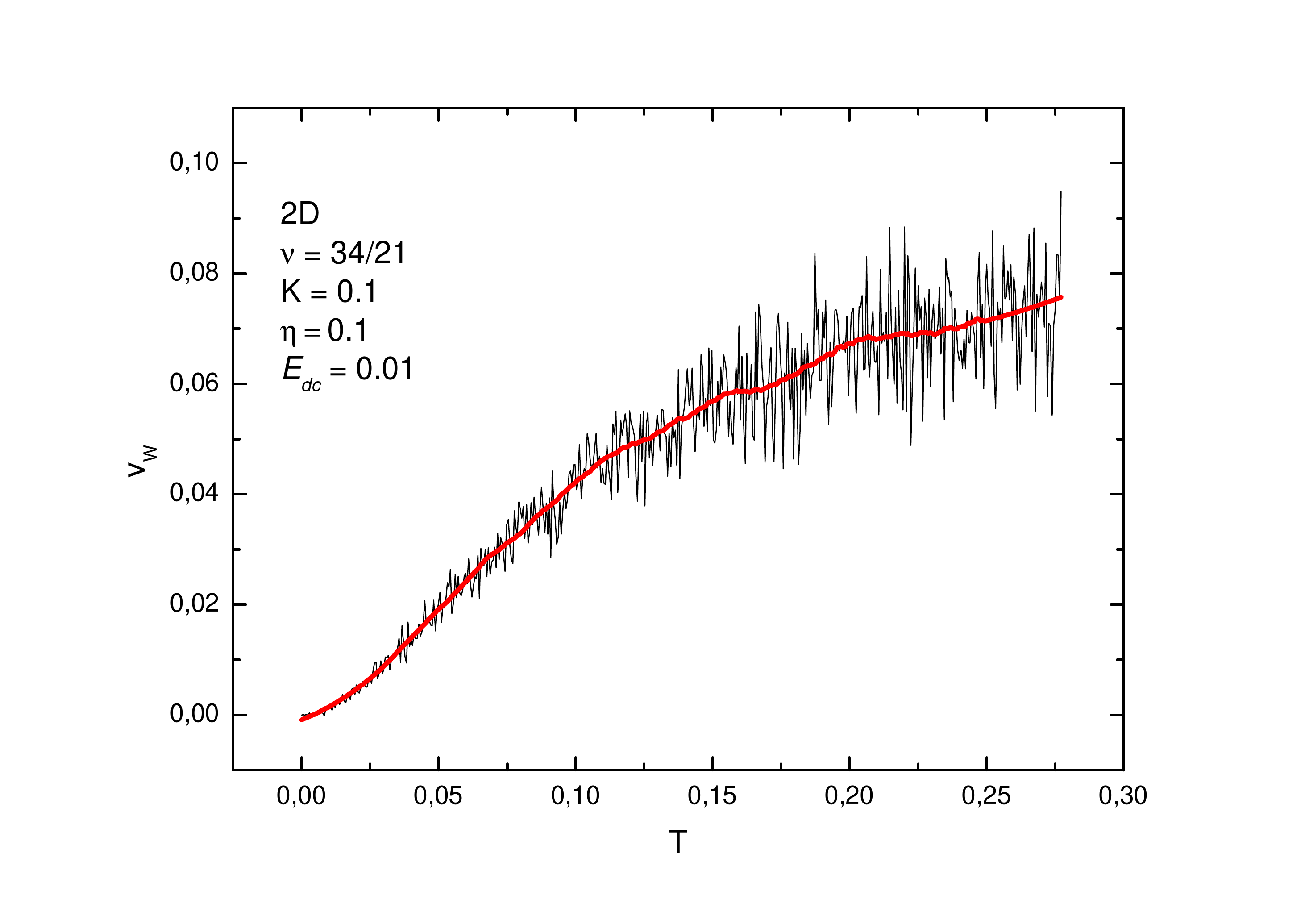}
\includegraphics[width=0.23\textwidth]{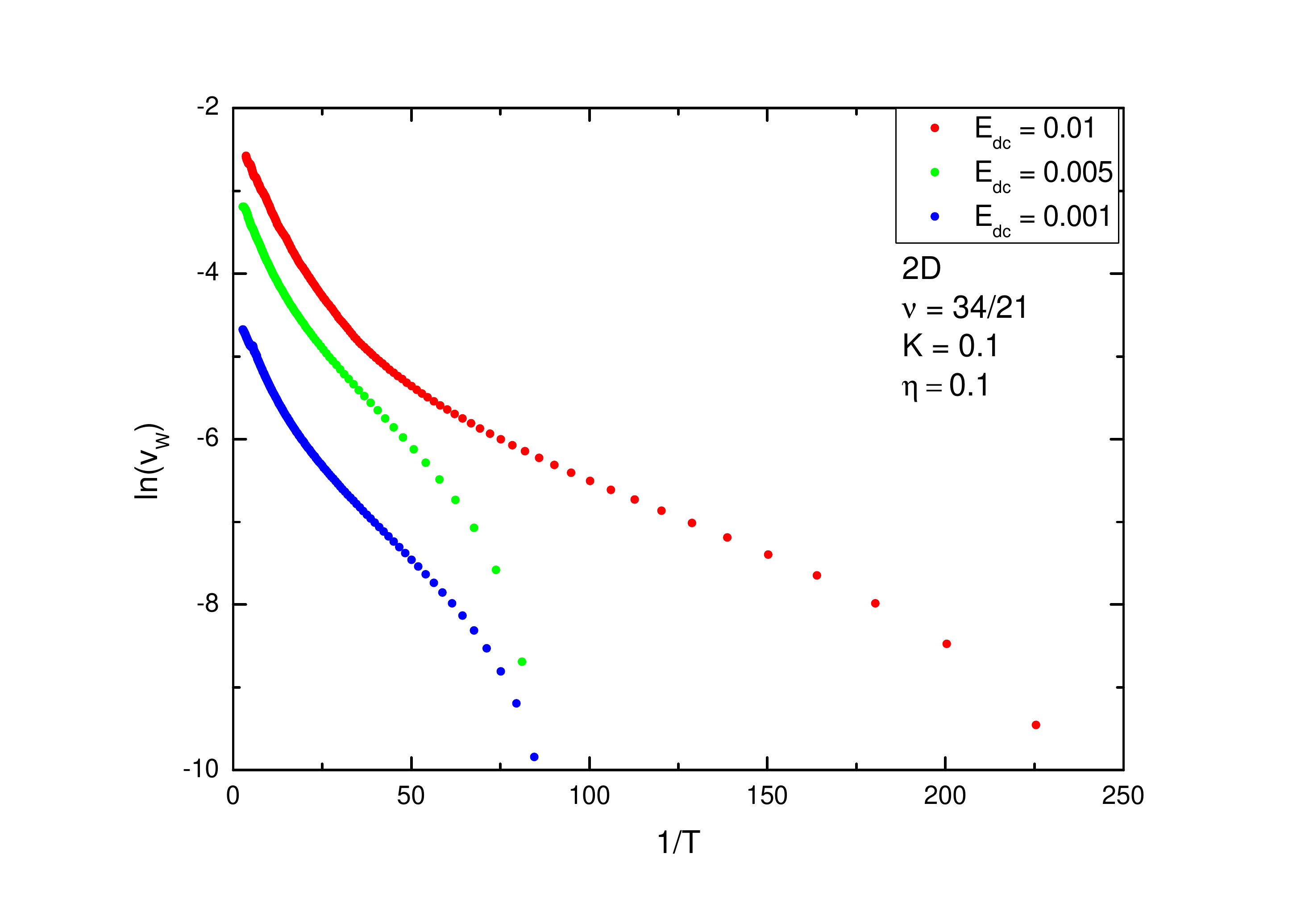}
\end{center}
\caption{\label{fig3} 
Left panel: dependence of Wigner crystal velocity 
$v_W$ on temperature $T$ at $K=0.1$, $L=21$, $L_y=5$,
$N_{tot}=21 \times 5$, $\nu=N_{tot}/LL_y = 34/21$ and 
and driving force $E_{dc} = 0.01$;
data are smoothed by Savitzky-Golay filter with polynomial order 2 
(points of window 150 in ORIGIN package)
shown by red curve.
Right panel: Thermal activation dependence of $ln(v_W)$ on inverse 
temperature  $1/T$ at parameters of left panel and 
different values of driving force $E_{dc}$, points show filtered data.
 }
\end{figure}

The thermoelectric propertied of the 2D system
are studied for typical values $K \geq 0.1$ and $T \sim 0.1$.
In such a regime a typical dependence of $v_W$ on $T$ is shown in 
Figure~\ref{fig3}. The obtained $v_W$ values are characterized
by a significant decrease of $v_W$ with decrease of $T$.
The presence of fluctuations can be overcome by
an averaging of data over Savitzky-Golay filter
showing that on average the data
are well described by the Arrhenius thermal activation dependence
$\ln v_W = -B - A_r/T$ which works for a large temperature range $T> 1/50$
(with $B = 4.67$, $A_R = 0.018$ at $E_{dc}=0.01$,
$ B = 3.74$, $A_R = 0.047$ at $E_{dc}=0.005$ and
$ B = 5.25$, $A_R = 0.044$ at $E_{dc}=0.001$).
The fit parameters show that 
for such $E_{dc}$ values
the current is described by the linear response
dependence $v_W \propto E_{dc}$.

Above we discussed the square-lattice case with $V=V_1$ in (\ref{eq:ham2}).
Similar results are obtained for two other lattices
with $V=V_2$ and $V=V_3$.
In the next Section we present the analysis of the 
thermoelectric properties in the linear response regime
for these three lattice geometries.

We note that the self-diffusion of electrons in 2D periodic potential
had been discussed recently in \cite{dykman}
but thermoelectricity and the Aubry pinned phase had not been analyzed there.

\section{Seebeck coefficient}
\label{sec:5}

To compute the Seebeck coefficient of our system we use the procedure developed in
\cite{ztzs}.
We use the Langevin description of a system evolution being a standard approach
for analysis of the system  when it has a fixed  
temperature created by the contact with the thermal bath or 
certain thermostat. The origins of this thermostat are not 
important since this description is universal \cite{politi}.
For the computation of the Seebeck coefficient $S$
we  create a temperature gradient along $x$-direction.
In the frame of the Langevin equation this is realized
easily simply by imposing  in (\ref{eq:langevin})
that $T$ is a function of an electron position along
$x$-axis with $T=T(x)=T_0 + G x $.
Here $T_0$ is the average temperature along the 
chain and $G=dT/dx$ is a small temperature gradient
(here $x$ is a coordinate position of a given electron).
For numerical computation of $S$ we have periodic conditions
in $y$-axis and we introduce  an elastic wall at $x=0$
keeping the Coulomb interactions
of electrons through this wall (this makes density
distribution homogeneous in absence of $E_{dc}$ and
temperature gradient).

\begin{figure}[t]
\begin{center}
\includegraphics[width=0.48\textwidth]{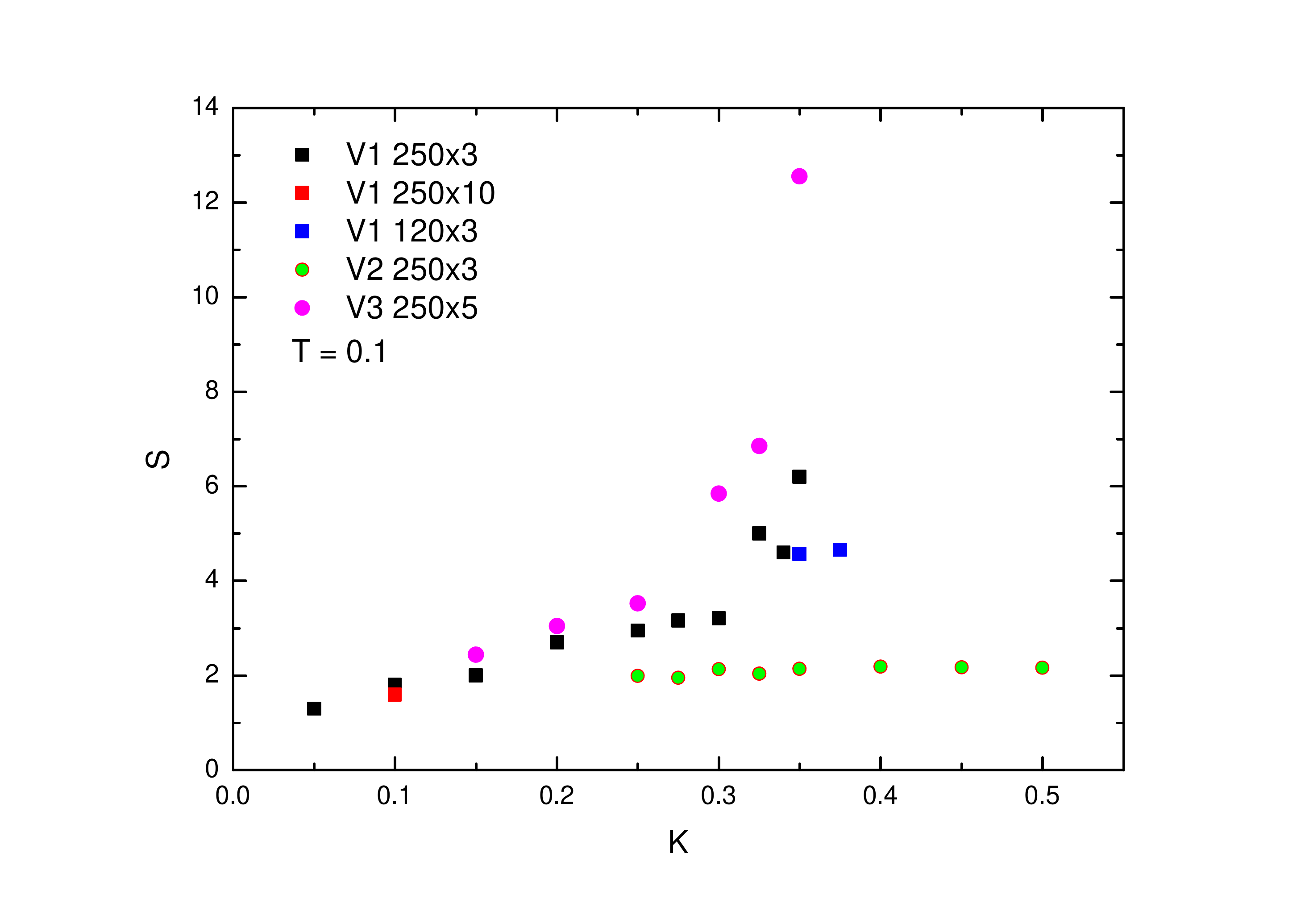}
\end{center}
\caption{\label{fig4} Dependence of Seebeck coefficient $S$ on
$K$ for different periodic potentials
given in (\ref{eq:ham2}) marked by $V_1, V_2, V_3$
with number of periodic cells $L_x, L_y$ given;
the total number of electrons being
$N \approx 1.618 L_x L_y$ and 
fixed temperature $T=0.1$ 
(here for $V=V_3$ case $K_h=0.002$).
 }
\end{figure}

At fixed temperature $T$
we apply a static field $E_{dc}$ which
creates a voltage drop 
$\Delta V =  2\pi L E_{dc} $ and
a gradient of electron density $\nu(x)$
along the chain. Then at $E_{dc}=0$ within
the Langevin equations (\ref{eq:langevin}) 
we impose a linear gradient of temperature
$\Delta T$  along $x$-axis and in the stabilized 
steady-state regime determine the electron density
gradient  $g(\nu) = d\nu(x)/dx$ along $x$-direction.
The data are obtained in the linear regime
of relatively small $E_{dc}$ and $\Delta T$ values.
Then the Seebeck coefficient
is computed as $S=\Delta V/\Delta T$
where $\Delta V$ and $\Delta T$
are taken  at such values
that the density gradient from $\Delta V$  
compensates those from $\Delta T$.
Examples of such density gradients
in presence of $E_{dc}$ and temperature gradient
are shown in Appendix (Figures~\ref{figA1},\ref{figA2},\ref{figA3},\ref{figA4}).

The obtained dependencies of $S$ on temperature $T$ at
different amplitudes $K$ of the periodic potential
are shown in Fig.~\ref{fig4} for all three geometries
of periodic potential given in (\ref{eq:ham2}). 
We discuss the dependence $S(K)$ for each geometry case.

For the square-lattice with $V=V_1$ in (\ref{eq:ham2})
we find a significant increase of $S$ with $K > K_{c2}$
at fixed temperature $T=0.1 >  K_{c2}$. At $K=0.1$ the values of $S$
are not affected by a variation of system size
from $L_y=3$ up to $L_y=10$. At largest value of $K=0.35$ we
obtain the largest value of $S= 6.2$ at $L_y=3$. 
Unfortunately, at such large $K$ values 
very long simulation times are required to reach
the steady-state in this strongly pinned Aubry phase.
We expect that at larger transverse size $L_y$ longer times 
are required to reach the steady-state and our maximal 
simulation time $t=2 \times 10^6$ is not sufficient
for $L_x=250$ and $L_y > 3$ for $K \approx 0.3$. A decrease of number of cells
in $x$-axis down to $L_x=120$ at $L_y=3$ leads to a moderate 
reduction of $S$ down to $S=4.57$ from its value $S= 6.2$ at $L_x=250$ 
at $K=0.35$. However, in this strongly pinned regime
we have rather strong fluctuations of $S$ with small variations of $K$
and we attribute this variation with $L_x$ to fluctuations.  
We checked that an increase of $T$ from $0.1$ up to $0.3$ at $K=0.1$
($L_x=250, L_y=3$)
leads to a reduction of $S$ approximately by $20\%$.
We checked that an increase of $T$ from $0.1$ up to $0.2$ at $K=0.3$ 
($V_3$, $L_x=250, L_y=5$) leads to a reduction of $S$ approximately by $14\%$.

From a physical view point at large values $T \gg K$
the influence of periodic potential becomes small
and we are getting moderate $S \sim 1$ values
corresponding to the sliding KAM phase. 
A similar dependence of $S$ on temperature 
has been found for 1D case (e.g. see right panel of Fig.3
in \cite{fki2007}).

The case of a diamond lattice with $V=V_2$ in (\ref{eq:ham2})
is also presented in Figure~\ref{fig4}.
In this case the dependence $S(K)$ is practically absent 
($S$ is increased only by 10\%  when 
$K$ is increased from $0.25$ up to $0.5$).
Thus the comparison of square and diamond lattices shows that
the lattice geometry place a significant role.
However, the reasons for significantly smaller $S$ values for the diamond
lattice remain not rather clear.
It is possible that for the diamond case there is a winding path that
keeps the sliding of electrons in this case 
so that it is more close to the sliding KAM regime
with moderate $S \sim 1$ values.

The strongest values up to $S \approx 12.5$  are found for
the case $V=V_3$ for $K_h=0.002$ (see Figure~\ref{fig4}). In this case the periodic
potential is only in $x$-direction while in $y$-direction we have
a harmonic potential. In a certain sense in this case
there are no any free path for flying through the system
at large values of $K$. Thus we assume that the pinned phase
is more robust for such a geometry. We note that the results
in Figure~\ref{fig4} are shown for $K_h=0.002$. At such a value of $K_h$
the harmonic potential is relatively weak and electrons still 
cover all $L_y=5$ cells in $y$-direction. 
As for the square lattice case we find that
$S$ is decreasing with an increase of $T$
(see Figure~\ref{figA5} of Appendix).

For $K=0.3$ we checked that an increase of $K_h$ from
$K_h=0.001$ up to $K_h=0.008$ leads to a decrease of
$S$ from $S=6.25$ down to $S=4.62$ 
(at fixed $T=0.1$ and $L_x=250, L_y=5$). We interpret this result 
assuming that at small $h$ values electrons have more flexibility
in $y$-direction that leads to larger $S$ values.
At $K_h=0.002$ we also checked that an increase of $L_y=5$
up to $L_y=11$ is sufficient to create a channel of 
electrons distributed in such a way that they
do not touch the boundary in $y$-direction;
however, in such a case we obtain $S=9.3$ 
(at $L_y=11$, $K_h=0.002$, $K = 0.3$) 
being higher compared to 
the case presented in Figure~\ref{fig4}
with $S=5.85$ at $L_y=5, K_h=0.002, K=0.3$.
We explain this by the fact that 
for $L_y=11$ the effective electron density is decreased
comparing to $L_y=5$ case (the total number of electorns
is the same in both cases) and thus 
the electron interactions are effectively reduced  
that leads to a more pinned regime with a larger $S$ value.

The results discussed above are obtained for a fixed 
electron density $\nu \approx 1.618$,
We checked that for $V=V_3$ case the value
of $S$ is not significantly affected by an increase
of $\nu$ up to $\nu =2.618$ where we 
obtained $S \approx 4$ at $K=0.3$, $L=250, L_y=5, K_h=0.006$. 
However, further more detailed investigation 
of dependence of $S$ on density $\nu$ are highly desirable.

The obtained results show that it is possible
to have rather large Seebeck coefficients
$S \approx 12 \gg 1$ at certain lattice geometries
in the Aubry pinned phase.

Of course, it would be very interesting to 
obtain the figure of merit $ZT$ for
the above lattices. However, the 
computation of thermal
conductivity $\kappa$, following the procedure described in 
\cite{ztzs,lagesepjd}, was not stabilized at maximal computational times 
$t=2 \times 10^6$. We attribute this to  long times
required for phonon (plasmon) equilibrium to be  reached
in our 2D system with about 2000 electrons.
Indeed, in 1D studies reported in \cite{ztzs,lagesepjd}
much larger times had been used ($t \sim 10^8$)
with smaller system sizes and about 50 - 100 electrons.

\section{Discussion}
\label{sec:7}

In this work we presented numerical modeling
of electron transport and themoelectricity
in 2D periodic lattices of different geometries.
We note that similar to 1D case discussed in
\cite{ztzs,lagesepjd}
there is a transition from sliding KAM phase
at $K < K_{c2}$ to the Aubry pinned phase
at  $K > K_{c2}$ where the electron current drops
exponentially with increase of $K$. 
However, compared to 1D case this transition 
is not so sharp probably due to presence of
more complex pathways for sliding of electrons.

While the KAM phase has moderate values of Seebeck coefficient 
$S \sim 1$ the Aubry phase is characterized 
by a significant growth of
$S$ with $K$ up to
the highest value $S \approx 12$ found in our numerical
simulations. At the same time it is established
that a change of geometry can lead to a significant
reduction of $S$ at the same amplitudes $K$ of periodic potential.
We attribute such a feature to presence of
free electron pathways crossing the whole system 
at certain geometries thus reducing 
maximal $S$ values. 

The maximal value of $S \sim 12$ obtained in this study
is still smaller than the extreme values
of $S$ obtained in certain experiments.
Thus in experiments with quasi-one-dimensional conductor
$\rm(TMTSF)_2PF_6$ \cite{espci} 
as high as $S =400 k_B/e \approx 35 m V/K$ value
 had been reached at low temperatures
(see Fig.~3 in \cite{espci}). Rather high 
values of $S \approx 50 k_B/e$ had been observed
in highly resistive two-dimensional semiconductor
(pinned) samples of micron size 
(see Fig.~8 in \cite{pepper}). The high values of $S \approx 30 k_B/e$
have been reported recently for CoSbS single crystals \cite{kotliar}.

In this studies we did not reach such high $S$
values but we obtain a clear dependence
showing that $S$ is rapidly growing with increase of 
potential amplitude $K$ and that it is also
growing with a decrease of temperature $T$.
Unfortunately, very high $S$ values appear only inside 
the strongly pinned Aubry phase where the times
of numerical simulations become very large 
to reach the steady-state regime. 
We are restricted by CPU time available
for our numerical simulations and thus we were not able
to penetrate inside such strongly pinned phase.
But our results clearly show that even higher value $S \gg 10$ 
can be reached in the strong pinned regime.

We think that the proposed investigations of themoelectric properties
of Wigner crystal in 2D periodic lattice   
are well accessible for experiments with 
low temperature electrons on a surface
of liquid helium in the regimes similar
to those discussed in \cite{kono1d,konstantinov}.
Indeed, for a typical lattice period
$\ell = 1 \mu m$ the potential
amplitude $K=0.1$ corresponds
to $V_A = K e^2/(\ell/2pi) \approx 10 K$ (Kelvin)
that can be reached at rather weak potential 
modulation in space. Such  $V_A$ can be
significantly larger than electron temperature
which easily takes values of $T = 0.1K$.
Thus we expect that such experiments will allow to obtain understanding
of fundamental properties of thermoelectricity.
As discussed in \cite{tosatti1} they can be also very useful 
for understanding of the fundamental aspects of friction
at nanoscale.

\section{ Acknowledgments}
 We thank N.~Beysengulov,
A.D.~Chepelianskii, J.~Lages, D.A.~Tayurskii and O.V.~Zhirov 
for useful remarks and discussions.

This work was supported in part by the Programme Investissements
d'Avenir ANR-11-IDEX-0002-02, reference ANR-10-LABX-0037-NEXT 
(project THETRACOM).
This work was granted access to the HPC GPU resources of 
CALMIP (Toulouse) under the allocation 2019-P0110.
The development of the VexCL library was partially 
funded by the state assignment
to the Joint supercomputer center of 
the Russian Academy of sciences for scientific
research. The work of M.Y. Zakharov was partially funded 
by the subsidy allocated to Kazan Federal
University for the state assignment in 
the sphere of scientific activities (project
$N^{\circ}$ 3.9779.2017/8.9).

\section{Author contribution statement}

All authors equally contributed to all stages of this work.

\clearpage
{\parindent=0cm \bf \large APPENDIX \vspace*{0.3cm}}

\setcounter{equation}{0}
\setcounter{section}{0}
\setcounter{table}{0}
\renewcommand{\theequation}{A.\arabic{equation}}
\setcounter{figure}{0}
\renewcommand{\thesection}{A.\arabic{section}}
\renewcommand\thefigure{A.\arabic{figure}}
\renewcommand\thetable{A.\arabic{table}}

\section{Appendix}

\begin{figure}[!h]
\centerline{\includegraphics[width=0.48\textwidth]{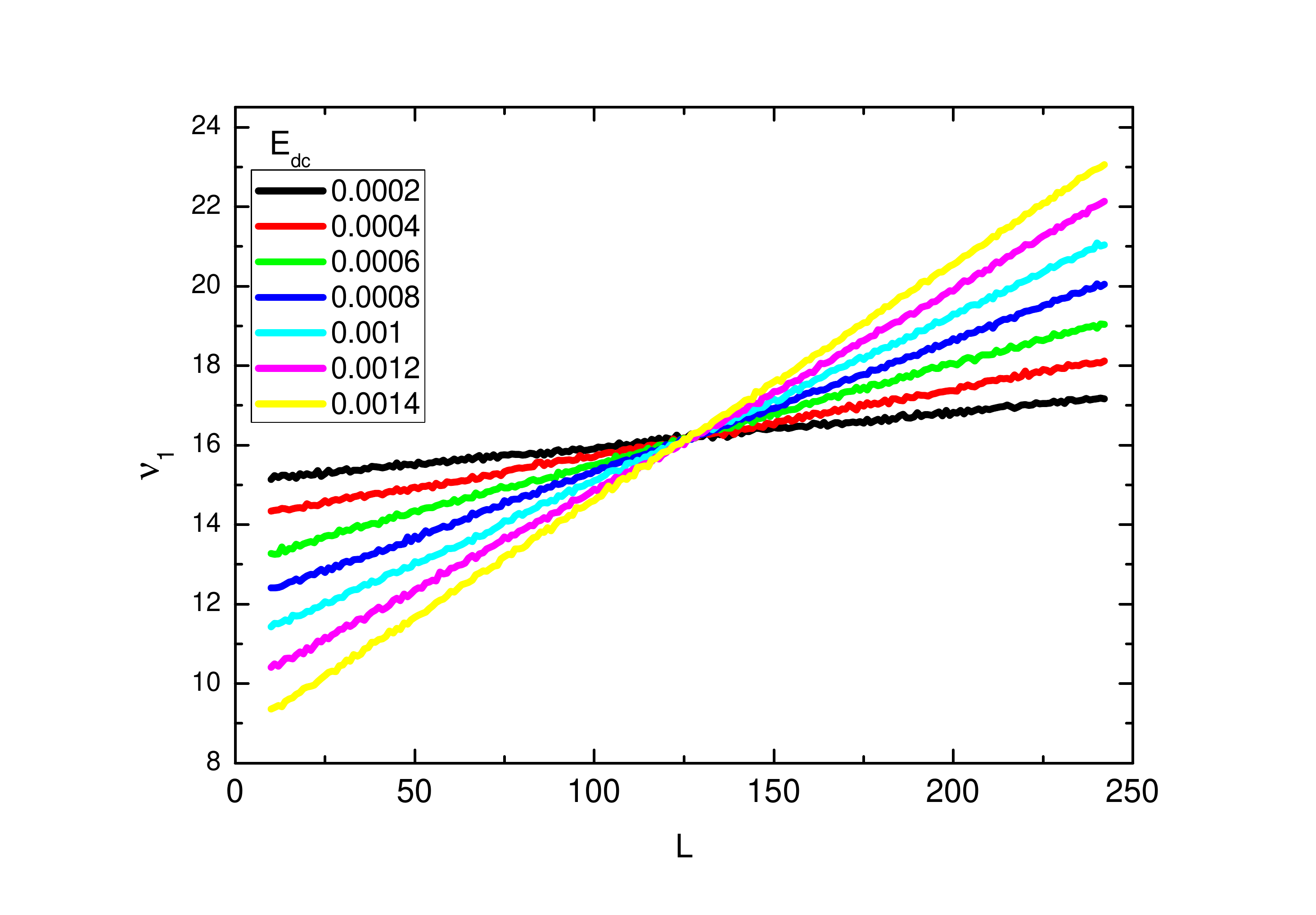}}
\caption{
Dependence of 1D electron density (averaged over $y$-axis) 
on $x=L$ coordinate at different values of applied static field 
$E_{dc}$. Here $V=V_1$ in (\ref{eq:ham2}),
$K=0.1, T_0=0.1$, the system has $L L_y = 250 \times 10$
period cells with average density of electrons per cell
being $\nu \approx 1.618$ (in total $N_{tot} = 4045$ electrons);
the physical time is $t= 2 \times 10^6$.
}
\label{figA1}
\end{figure}

\begin{figure}[!h]
\centerline{\includegraphics[width=0.48\textwidth]{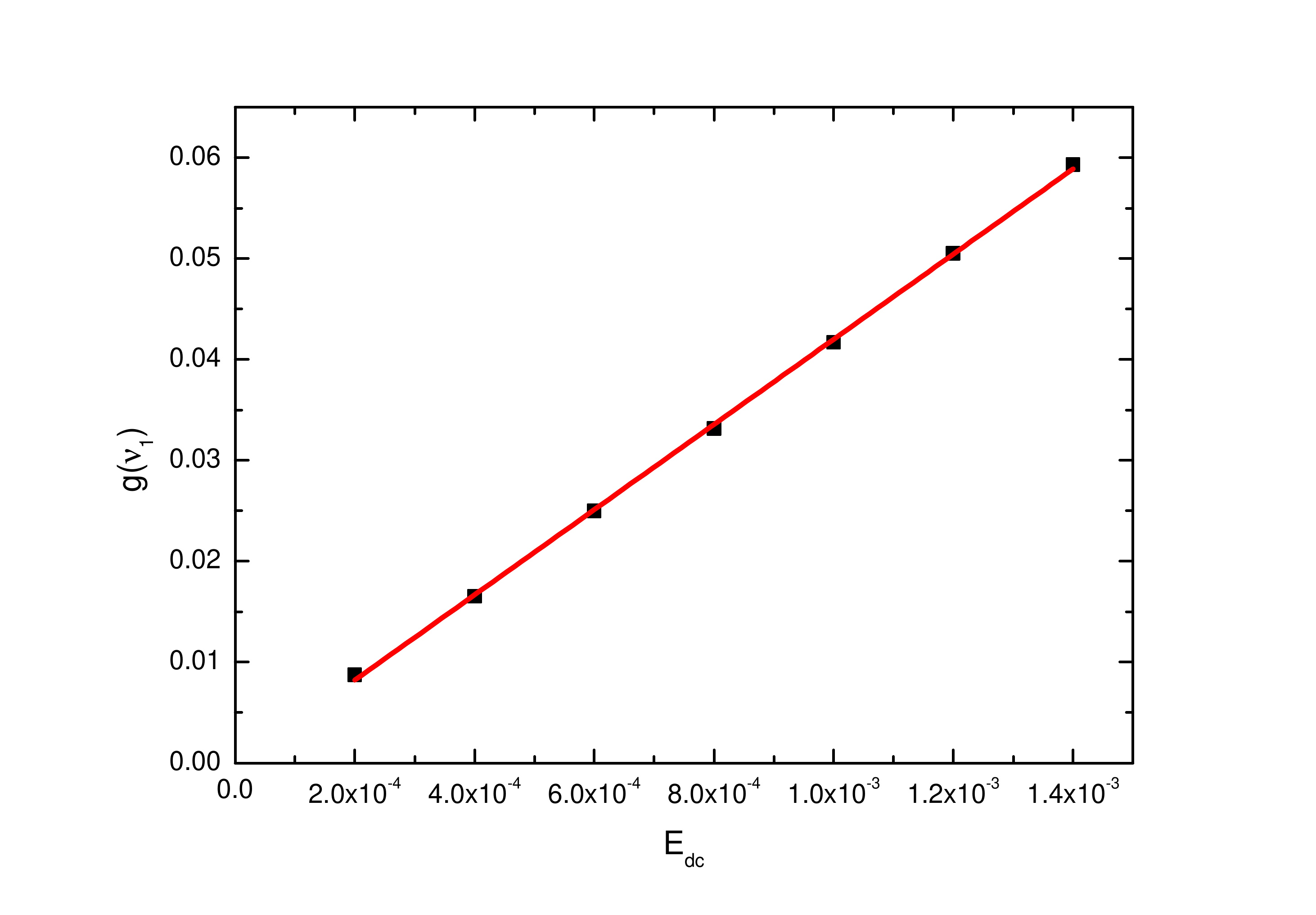}}
\caption{
Dependence of the gradient of electron density $g(\nu_1)$ 
on static  field $E_{dc}$ for parameters of Figure~\ref{figA1};
$\nu = \nu_1=1.618$.
}
\label{figA2}
\end{figure}

\begin{figure}[!h]
\centerline{\includegraphics[width=0.48\textwidth]{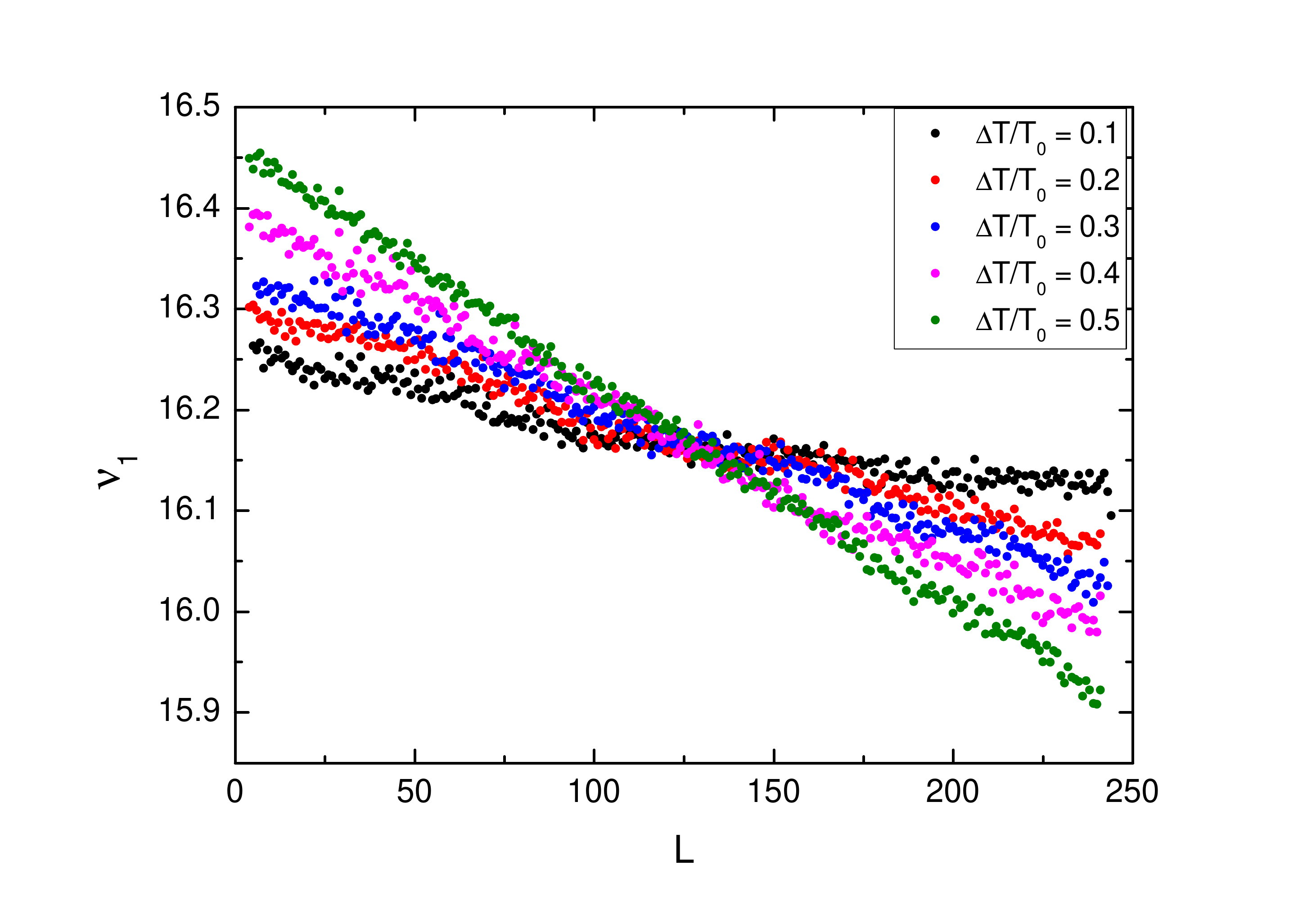}}
\caption{
Dependence of 1D electron density (averaged over $y$-axis) 
on $x=L$ coordinate at different values 
of temperature difference $\Delta T$ at the end of the sample;
other parameters are as in Figure~\ref{figA1}, $E_{dc}=0$. 
}
\label{figA3}
\end{figure}

\begin{figure}[!h]
\centerline{\includegraphics[width=0.48\textwidth]{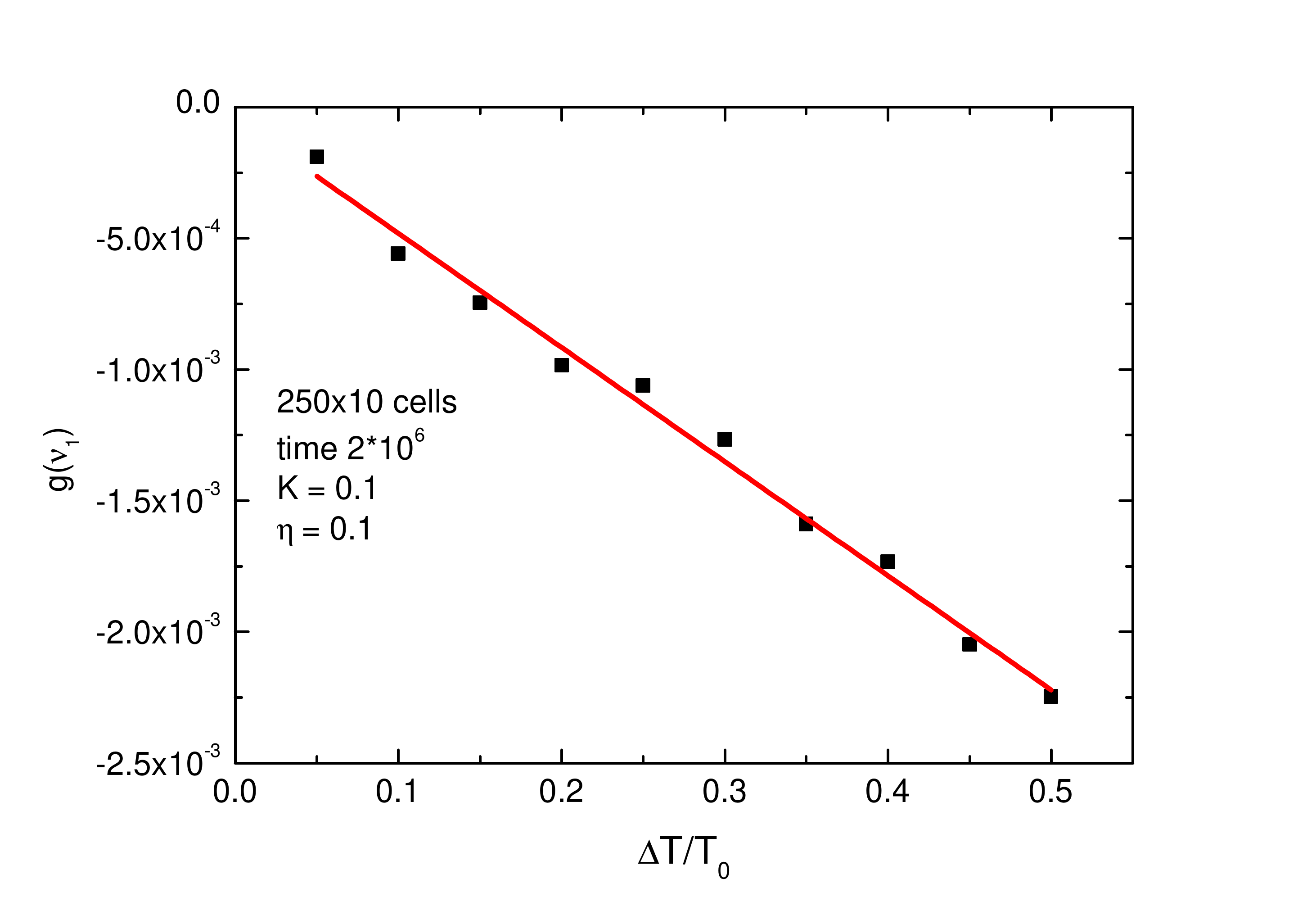}}
\caption{
Dependence of the gradient of electron density $g(\nu_1)$ 
on  temperature difference $\Delta T$ at the end of the sample;
other parameters are as in Figure~\ref{figA1}, $E_{dc}=0$. 
}
\label{figA4}
\end{figure}

In Figures~\ref{figA1},\ref{figA2},\ref{figA3},\ref{figA4}
we show the variation of electron density
induced by an external static field $E_{dc}$ and temperature gradient
$dT/dx \propto \Delta T/T_0$ (here $T_0$ is the average sample temperature
and $\Delta T$ is the temperature difference at the ends of the sample).

\begin{figure}[!h]
\begin{center}
\includegraphics[width=0.48\textwidth]{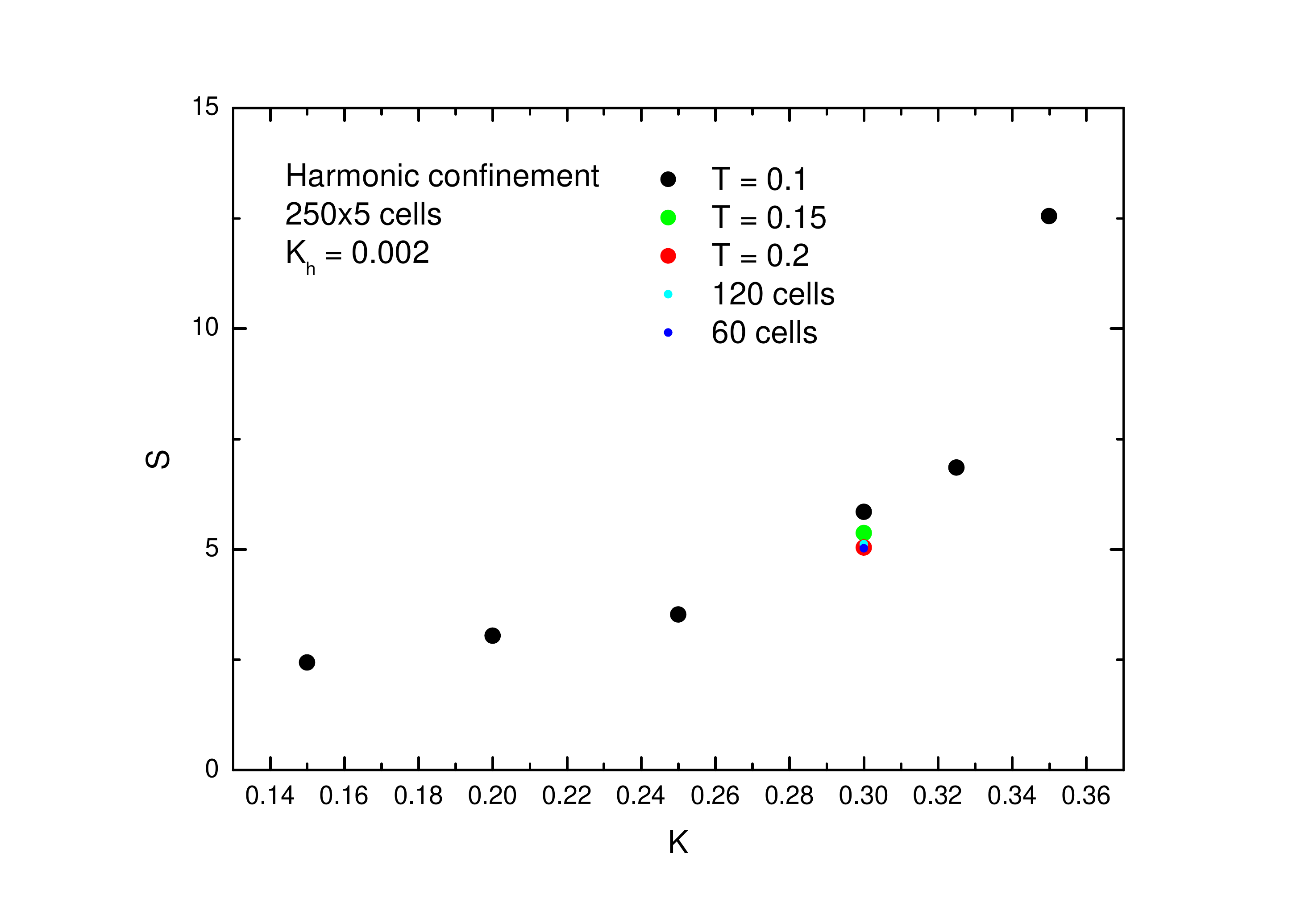}
\end{center}
\caption{
Dependence of Seebeck coefficient
$S$ on $K$
the system with harmonic confinement $V=V_3$
in (\ref{eq:ham2}) at $K_h=0.002$
for $L=250, L_y=5$ and $T=0.1$ (black points),
$T=0.15$ (green point), $T=0.2$ (red point);
cyan and blue points show data at $T=0.1$  for 
$L=120, L_y=5$ and $L=60, L_y=5$ respectively.
.
 }
\label{figA5}
\end{figure}

Figure~\ref{figA5} shows the dependence $S(K)$
for the harmonic channel at different temperature and system size
values.

\end{document}